\documentclass[11pt, showkeys]{revtex4}

\usepackage{amsmath,amsfonts,amssymb,amscd, euscript}

\usepackage[matrix,arrow,curve]{xy}

\usepackage[dvips]{graphicx}
\usepackage{epsfig}
\usepackage[normalem]{ulem}
\usepackage[utf8x]{inputenc}

\usepackage[T2A,T1]{fontenc}
\usepackage[russian,english]{babel}

\newtheorem{theo}{Theorem}[section]

\newtheorem{prop}{Proposition}[section]

\newtheorem{rem}{Remark}[section]
\newtheorem{dfn}{Definition}[section]
\newtheorem{ex}{Example}[section]

\def\rd{\mathrm{ d}}

\def\ud{\mathrm{ d}}

\newcommand {\supplus}{\mathop{{\supset}\llap{\raise
0.5pt\hbox{\normalfont\small+}\hskip 0.5pt}}}

\newcommand {\subplus}{\mathop{{\subset}\llap{\raise
0.5pt\hbox{\normalfont\small+}\hskip 0.5pt}}}

\newcommand {\suptimes}{\mathop{{\supset}\llap{\raise
0.5pt\hbox{\normalfont\small $\times$}\hskip 0.5pt}}}

\newcommand {\subtimes}{\mathop{{\subset}\llap{\raise
0.5pt\hbox{\normalfont\small $\times$}\hskip 0.5pt}}}

\def\Z{\mathbb{Z}}
\def\R{\mathbb{R}}

\begin{document}

\title{Supersymmetrization: AKSZ and beyond?}

\author{Vladimir Salnikov}
\email{vladimir.salnikov@uni.lu}

\affiliation{RMATH, University of Luxembourg,  \\
6, rue Richard Coudenhove-Kalergi
L-1359 Luxembourg}


\begin{abstract}
\noindent
In this paper we describe multigraded generalizations of some constructions useful for mathematical 
understanding of gauge theories:
we perform a near-at-hand generalization of the Aleksandrov--Kontsevich--Schwarz--Zaboronsky
procedure, we also extend the formalism of $Q$-bundles introduced first by A.~Kotov and T.~Strobl. 
We compare these approaches studying 
some supersymmetric sigma models important in theoretical physics.

\end{abstract}

\pacs{} \keywords{AKSZ procedure, Q-bundles, multigraded geometry, 
supersymmetrization, \newline 
graded Poisson sigma model}

\maketitle

\section{Introduction / motivation}

In this paper we present a couple of closely related mathematical approaches to the problem of supersymmetrization, coming from 
theoretical physics. The main motivation is to provide a convenient framework for studying the 
geometry behind gauge theories, and in particular supersymmetric ones. 
In the recent years there appeared several papers written by physicists where some theories 
are studied through lengthy ``by hand'' computations. 
We think that most of those can fit to a general framework heavily relying on multigraded geometry.

The main object of analysis will be \emph{sigma models} --- very informally speaking those are 
functionals between source (world-sheet) and target manifolds equipped with some geometry. 
Phrased like this, it is a very general and flexible framework to study gauge theories. 
The main question is what geometric structures provide convenient tools for reflecting the physics 
of the problem.
We will present a near at hand generalization of the classical Aleksandrov--Kontsevich--Schwarz--Zaboronsky
procedure (\cite{AKSZ}) to multigraded manifolds used as source and target. We will 
also see the multigraded version of the approach of \cite{AK-TS}
to studying gauge theories using $Q$-bundles and $Q$-morphisms. 
Main examples we give, are related to super Poisson sigma models and super Chern-Simons theory. 
They show that even though the $Q$-bundle approach formally permits 
to reproduce the AKSZ scheme, using both of them can give an enlightening description 
from various perspectives, depending on what questions one asks oneself.  

The paper is organized as follows: 
The next section is devoted to basic definitions from multigraded geometry, it is backed up by the appendix
\ref{sec:super_appendix}, where 
classical notions from super/graded geometry are sketched. Section \ref{sec:superaksz} is a straightforward generalization 
of the classical AKSZ procedure, we pay special attention to analysis of admissible 
structures on the source and the target. In section \ref{sec:superCW} we briefly sketch 
the $Q$-bundle approach pointing out key differences related to several gradings.
Various applications of these approaches are presented in section \ref{sec:susy}, namely 
we see whether adding super degrees of freedom to the source or the target lead to 
non-trivial supersymmetrization of given theories.

\section{Multigraded geometry} \label{sec:Gman}

In this section we set up the definitions related to multigraded manifolds -- they will be necessary 
both for the generalized AKSZ construction and for the $Q$-bundle approach. 
For this, we combine the notions of super- and graded manifolds 
to define all the necessary geometric data on multigraded manifolds and discuss various compatibility 
questions. For the sake of completeness (and to fix the notations) the classical notions from super and graded geometry are recalled 
in the Appendix \ref{sec:super_appendix}.

\subsection{Multigraded manifold}
The definition of a graded manifold given in the appendix can be naturally 
generalized to several gradings following the idea of \cite{GR11}.
Consider a manifold equipped with two commuting homogeneity structures $h^1$ and $h^2$, 
that is the corresponding homothety operators that satisfy 
$$ h^1_t \circ h^2_u  = h^2_u \circ h^1_t. $$
The coordinate transformations should preserve the eigenspaces of these operators.
Such a manifold will be called \emph{bigraded}.
The data of two homogeneity structures gives rise to the following diagram 
of graded manifolds
$$
\xymatrix{
  E   \ar [rr]^{h^2}  \ar[dd]_{h^1}     &&  E_2 \ar[dd]^{h^2_0} \\
&&&\\
  E_1  \ar [rr]^{h^1_0}&& M & 
}
$$
where $h^i_0$ are the restrictions of the corresponding homotheties.
This construction corresponds to the existence of two commuting Euler vector fields, 
one then can choose a coordinate system which will correspond to eigendirections 
of both maps, that is homogeneous with respect to the double $\Z \times \Z$ grading.
In the same way one can define multiple $\Z$-grading.

In what follows we will be interested in defining a slightly  
different case of a bigraded manifold being
a ``graded manifold over a supermanifold''. We make this distinction between 
$\Z_2$ and $\Z$ gradings voluntarily, to stress the fact that they are independent 
and in the applications have different ``nature'': $\Z_2$-grading comes from the \emph{physics}
of the problem and is related to description of bosonic vs fermionic degrees of freedom, 
while the $\Z$ ones are parts of the \emph{geometric machinery} we use to efficiently describe the 
problem.
The construction is morally similar to the one described in the previous paragraph:
the $\Z$-graded will be encoded in an Euler vector field (or equivalently 
homogeneity structure), 
with the only requirement that it should be  even  with respect 
to the $\Z_2$-parity coming from the super grading of the manifold. 
This is the analog of the requirement of commutation between 
the homogeneity structures. 

Given a parity even Euler vector field, one can again choose the 
bi-homogeneous (w.r.t. $\Z \times \Z_2$-grading) 
coordinate system and view the resulting bigraded manifold 
as a supermanifold 
$(M^{n|m}, {\cal O}_M)$  (the respective ($\Z_2$) parity will be denoted 
by $p(\cdot)$) with an extra $\Z$-grading on the level of the 
sheaf of functions (denoted by $gh(\cdot)$).
For a fixed coordinate system $x^i, \, i = 1, \dots, n+m$
the Euler vector field reads
$\epsilon = gh(x^i)x^i\frac{\partial}{\partial x^i}$. 
The modified structure sheaf can now be viewed as
${\cal O} = \oplus_i{\cal O}^i$, such that for any 
homogeneous function $f$ -- subsection of ${\cal O}^i$ $gh(f) = i$.

The algebraic structure is then defined in a similar way 
as in the ordinary $\Z$-graded case with the only remark 
that for all operations one needs to keep trace simultaneously 
of the parity $p(\cdot)$ and of the grading $gh(\cdot)$.
Here one should note, that there are two possible sign conventions when treating 
the bigraded objects:	
\newline \textbf{Bernstein -- Leites}: 
$$ 
x^\alpha \cdot x^\beta = (-1)^{(gh(x^\alpha) + p(\alpha))(gh(x^\beta) + p(\beta))}x^\beta \cdot x^\alpha 
$$ 
and \textbf{Deligne}:
$$ 
x^\alpha \cdot x^\beta = (-1)^{gh(x^\alpha)gh(x^\beta) + p(\alpha) p(\beta)} x^\beta \cdot x^\alpha 
$$
These two conventions are equivalent in a sense that there is an isomorphism (\cite{Deligne}) 
from an algebraic structure defined by one of them to the algebraic structure 
defined by the other. 
The definition of \emph{bigraded} given above is more suitable 
for the Bernstein--Leites convention, but in the physics literature the Deligne one 
is also  used. For the latter one the bigraded manifolds are defined via this isomorphism. 
In what follows we will prefer the Bernstein-Leites sign convention and also sometimes 
instead of $p(\cdot)$ we will
keep track of the total parity $par(\cdot) \equiv (p(\cdot) + gh(\cdot))\!\mod  2$.


The direct generalization of this construction is the multigraded manifold, 
when one assigns the grading to any coordinate: 
$gr (x^i) = (p^i_1, \dots, p^i_k,gh^i_1, \dots, gh^i_l) \in \Z_2^k\times\Z^l$.
Both sign conventions are easily generalizable to multiple gradings: 
in the former, one just takes the sum of all the degrees, no matter if they are 
$\Z$ or $\Z_2$; and in the latter the scalar product of two lines of degrees. 
In principal between various couples of the  $\Z$ and $\Z_2$
in $gr(x)$ there could be different sign conventions, but 
given a homogeneous function $f(x)$ it is important to be able to assign some 
total ghost grading $gh(f) \in \Z$ to it and distinguish 
whether the element is \emph{even} (i.e. commuting with anything)
or \emph{odd} (two such elements anticommute) -- this distinction 
is done naturally if all the conventions are Bernstein--Leites, otherwise 
the above mentioned isomorphism should be used. 

This definition is already enough for applications, although one can extend it 
to the case of $\mathrm{G}$-graded objects, $\mathrm{G}$
being an abelian semigroup 
equipped with two semigroup 
homomorphism $par:\mathrm{G}\rightarrow \mathbb{Z}_{2}$, called parity, and 
$gh:\mathrm{G}\rightarrow \mathbb{N}$ 
called ghost number. 
To recover the previous construction consider $\mathrm{G}=\Z \times \mathbb{Z}_{2}$ 
with $gh(n, \alpha)=n$, and 
$par(n, \alpha)=(n+\alpha)\!\mod 2$.
That corresponds to a bigraded manifold defined above
with the Bernstein-Leites sign convention for commutation relations. 


The space of \emph{homomorphisms} between two multigraded manifolds 
is also naturally multigraded.  
As in the $\Z$-graded case the space of 
homomorphisms between two vector spaces is defined by 
$\mathrm{\underline{Hom}}(V,W)=\bigoplus\limits_{g\in \mathrm{G}}\mathrm{Hom}_{g}(V,W)$ 
or
$\mathrm{\underline{Hom}}(V,W)=\bigoplus\limits_{g\in \Z_2^k\times\Z^l}\mathrm{Hom}_{g}(V,W)$ 
where $\mathrm{Hom}_{g}(V,W)$ is the space of all linear homomorphisms of degree $g$, 
$f:V\rightarrow W,\;\;f(V^{k})=W^{k+g}$. 
A \emph{morphism} is then a homomorphism
preserving the total degree ($g \in \mathrm{G}$ or $g \in \Z_2^k\times\Z^l$ 
depending on the setting). 
We will however be sometimes interested in the 
homomorphisms preserving only some degrees, so this notion is more subtle.

\begin{rem}
In principal one can just consider $G = \Z\times \Z_2$ from the very beginning 
when looking at one $G$-graded manifold. The only subtlety is that when one considers a morphism between
two $G$-graded manifolds their parity parts need not be coherent, it means that 
to have a well defined category one needs to incorporate both parities into $G$
and distinguish the manifolds by the parity morphisms.
The construction certainly remains valid if the $\Z_2$ parity of all the $G$-graded manifolds 
in the category coincides. 
\end{rem}
\begin{ex} Let us consider the $\mathbb{N}\times \mathbb{Z}_{2}$ product manifold 
$T[1]\Sigma\times \mathbb{R}^{p|q}$ ($\mathbb{N}$ means $\Z_{\ge0}$ here); 
$T[1]\Sigma$ is a standard $\mathbb{N}$-graded manifold and 
$\mathbb{R}^{p|q}$ a $\mathbb{Z}_{2}$ superspace. 
If $\mathcal{C}^{\infty}(U)\otimes \Lambda ({\theta_{1},...,\theta_{n}})$ is the structure sheaf of 
$T[1]\Sigma$ and $\mathcal{C}^{\infty}(\mathbb{R})^{p}\otimes \Lambda ({\xi_{1},...,\xi_{q}})$ the corresponding one 
for $\mathbb{R}^{p|q}$, the parity and ghost morphisms are given by:
$$
par \colon f_{i_{1}..i_{k}j_{1}..j_{l}}\theta^{i_{1}}...\theta^{i_{k}}\xi^{j_{1}}...\xi^{j_{l}}
\mapsto (k+l)\mathrm{mod}2
$$
$$
gh \colon f_{i_{1}..i_{k}j_{1}..j_{l}}\theta^{i_{1}}...\theta^{i_{k}}\xi^{j_{1}}...\xi^{j_{l}}\mapsto k.
$$
\end{ex}
\begin{rem}
  In general the space of functions on  multigraded and even just on  $\Z$-graded manifolds 
  can present some difficulties from the functional analytic point of view, see the discussions 
  in \cite{JPS, KwPo, CoPo}.

\end{rem}

By a direct analogy with the section \ref{sec:gman} the multigraded manifolds form 
a category ${\cal MG}Man$ with $Hom(M,N) = Hom({\cal O}_N, {\cal O}_M)$.
Two forgetful functors are defined: into ${\cal G}Man$ (keeping only the $gh$-grading)
and into ${\cal S}Man$ (keeping only the $par$-parity); in both cases in the 
target category smooth functions of the even (degree 0) variables are 
added to the sheaf of functions. 
 The space of maps between two multigraded manifolds $M$ and $N$ can be functorially defined 
 similarly to the proposition (\ref{prop_gm}).


\subsection{$Q$ and $P$ structures on multigraded manifolds}

\begin{dfn}
  A \emph{bigraded $Q$-manifold} is a multigraded manifold endowed with a \emph{$Q$-structure} 
  -- a ghost degree $gh = 1$ par-parity odd  homological (i.e. self-super-commuting) vector field.
\end{dfn}

Typical examples in this paper will be produced from the following  manifolds: \newline 
$(\Sigma^{n|m}, d_{\text{de Rham on }\Sigma})$ -- supermanifold with the de Rham differential acting 
only on the even part; \newline 
$({\cal G}[1], d_{CE})$ -- super Lie algebra with the super Chevalley-Eilenberg 
differential, \newline 
$(T^*[1]M, Q_{\pi})$ -- shifted cotangent bundle associated to a super Poisson manifold.  
 
As in the standard  setting, a symplectic structure (a non-degenerate closed two-form) 
$\omega~=~dx^{i}\omega_{ij}dx^{j}$ is \emph{even} 
if $par(\omega_{ij})=par(i)+par(j)$ and \emph{odd}
if $par(\omega_{ij})=par(i)+par(j)+1$.

\begin{dfn}
  A \emph{$P$-structure} is a ghost degree $gh = 1$, par-parity even symplectic structure on a 
  multigraded manifold.
\end{dfn}
A typical example will be a degree $1$, par-parity even form $\omega = \rd p_i \rd x^i$ 
canonically associated to $\Pi T^*M$ (or $T^*[1]M$), $M$ being a supermanifold. 

\begin{rem}
 For multigraded manifolds with several $\Z$ gradings we will systematically specify 
 which combination of ghost degrees should be equal to $1$.
\end{rem}

As in the ordinary ($\Z$-graded) case a symplectic structure defines a bracket of two functions
$(f,g) := \iota_{X_f}\iota_{X_g}\omega$, where $X_f$ is the Hamiltonian vector field of $f$, 
that satisfies $\iota_{X_f} \omega =  \rd f$.
For the case of a $P$-structure this bracket is the analog of a Gerstenhaber or BV bracket. 
For a degree $0$ parity even symplectic structure it is a super Poisson bracket. 

A vector field $X$ is \emph{compatible} with the symplectic form 
if ${\cal L}_X \omega = 0$, where the usual notion of a Lie derivative extends to 
multigraded objects by the graded Cartan's formula: \newline
${\cal L}_X = \iota_X \rd + (-1)^{par(X)} \rd \iota_X$.
\begin{rem}
  In general the definition of a Lie derivative should be compatible with the 
  chosen sign convention. Namely, the sign rule is $v\otimes w = (-1)^{sign(v, w)} w\otimes v$
  permits to define
  ${\cal L}_X := [\iota_X, d]_{sign}$, where $[\cdot, \cdot]$ is the graded commutator governed 
  by $sign(\cdot, \cdot)$.  In the case of a multigraded manifold with the 
  Bernstein--Leites sign convention, and $par(\cdot)$ being the sum of all degrees, 
  one recovers the above formula for ${\cal L}_X$, which reduces to the 
  usual Cartan's formula in the ungraded case.

\end{rem}

\begin{dfn}
  A multigraded $QP$-manifold is a multigraded manifold with compatible $Q$ and $P$ structures. 
\end{dfn}
If $Q$ is Hamiltonian, its Hamiltonian function ${\cal Q}$ is of ghost degree $0$ and of even parity, 
and satisfies $({\cal Q},{\cal Q})=0$ (master equation). 

\subsection{Integration}
\begin{dfn} 
  A \emph{partial measure} $\mu$ on a multigraded manifold $M$ is a linear functional on the algebra of 
  all functions
  on $M$ taking values in the $gh=0$-graded functions (i.e. $gh( \int \mu f ) = 0$). 
  A partial measure is \emph{homogeneous} of degree $(-n)$ if
  for homogeneous elements
    $gh( \int \mu f ) = gh(f) - n$. 
  A partial measure is called \emph{non-degenerate} if it defines a non-degenerate bilinear  form 
on the algebra of functions. Given a vector field $Q$, the measure $\mu$ is \emph{$Q$-invariant} if
 $\int \mu Q f = 0, \, \forall f$. 
The parity of the defined measure depends on the number of odd variables integrated out.
    A \emph{full measure} on a bigraded (multigraded) manifold is a partial measure which is real-valued 
(or valued in an auxiliary Grasmann algebra). 
\end{dfn}
\begin{ex} In the case of  $M = T[1]\Sigma\times\R^{p|q}$ for $\Sigma$ an orientable manifold with canonical 
measure $vol$ and the associated berezinian $\mathrm{Ber}$ on $T[1]\Sigma$; if 
$\rho$ the berezinian on $\mathbb{R}^{p|q}$ the total measure is just given by the tensor product, 
i.e. $\mu = \mathrm{Ber}\otimes \rho$.
\end{ex}

\begin{rem}
Constructing homogeneous non-degenerate measures on generic multigraded manifolds is not an easy task. 
We shall comment on the possible obstructions below after having defined the multigraded
AKSZ construction. 
\end{rem}

\section{Multigraded AKSZ}
In this section we will assemble the previous definitions to formulate the ``multigraded'' analogue of the 
AKSZ construction (\cite{AKSZ}).

\label{sec:superaksz} 

\subsection{$QP$-structure on $Y^X$}

\begin{theo}\label{gaksz}

Consider two multigraded manifolds $X$ (source) and $Y$ (target).
Let the source $X$ be equipped with a $Q$-structure $D$ and a $D$-invariant 
homogeneous ($gh$-degree $-(n+1)$) non-degenerate measure $\mu$; 
the target $Y$ equipped with a $Q$-structure $Q$, compatible with the symplectic structure 
$\omega$, such that $gh(\omega) = n$ and the parity is opposite to the parity of $\mu$.
Then the space of (multigraded) maps $Y^X$ can be equipped with a $QP$-structure.
Moreover if $\omega$ is exact one can define a functional on $Y^X$ 
satisfying the classical master equation. 
\end{theo}
\textbf{Proof.} We construct the $QP$ structure on the space
$Y^X$ following the scheme reviewed in \cite{CatFel}.

\textbf{$Q$-structure.} Define $\tilde Q = \hat D + \check Q$, where 
$\hat D (x, f) = \rd f(x)D(x)$, $\check Q(x,f) = Q(f(x))$. Since 
the superdiffeomorphisms of source and target induce respectively left and right 
actions on $Y^X$, they supercommute. That is $\tilde Q$ is a self-super-commuting vector field  
of appropriate multidegree as $\hat D$ and $\check Q$ are $Q$-structures by themselves.

\textbf{$P$-structure.}
Consider the evaluation map $ev \colon X\times Y^X \to Y, \quad (x,f)\mapsto f(x)$.
This permits to define a $P$-structure (of ghost degree 1) on $Y^X$: $\tilde \omega = \mu_* ev^* \omega$. 
$\tilde \omega$ is clearly closed, its non-degeneracy and closedness are guaranteed by 
the properties of $\mu$ and $\omega$. 

\textbf{Compatibility.}
With the same argument as 
in \cite{CatFel} we can show that if the hamiltonian function of $Q$ with respect to $\omega$
is $S$, then the hamiltonian function of $\check Q$ with respect to $\tilde \omega$
is $ \check S = \mu_* ev^* S$, that is $(\check Q, \tilde \omega)$ is a $QP$-structure.
Since $\mu$ is $D$-invariant, ${\cal L}_{\hat D} \mu_* ev^* = 0$. 
It means that $(\tilde Q, \tilde \omega)$ equips $Y^X$ with a $QP$-structure.
Moreover, if $\omega = \rd\tau$, then $\hat D$ is hamiltonian, $\hat S = -\iota_{\hat D} (\mu_* ev^* \tau)$.
And for hamiltonian $Q$ one can define $\tilde S = \check S + \hat S$, which satisfies the 
master equation $(\tilde S, \tilde S)_{\tilde \omega}$, where $( \cdot , \cdot )_{\tilde \omega}$ 
is the Gerstenhaber bracket on $Y^X$ associated to $\tilde \omega$. 

\begin{rem} 
Although technically the construction is rather similar to the classical 
(not multigraded)
case this simplicity is a bit misleading. 
The essential difference is in the definitions of admissible ingredients (measure, 
symplectic structure, 
compatibility etc.) since we are working in a larger category of multigraded manifolds.
Even the existence of those structures for some natural examples is not a trivial statement --
we shall be more precise on this in the following two subsections.
\end{rem}

\subsection{Target: Classification of supersymplectic structures.}

Note that for the arbitrary \textbf{super target}, that is when $M$ is 
bigraded: $\mathbb{Z} \times \mathbb{Z}_2$,
or more generally multigraded,  since the 
$Q$-structure influences only the ghost grading,
the procedure above doesn't produce any difficulties as soon as 
the notion of $Q$-morphisms between two multigraded manifolds is well defined (see also \cite{CatFel}). 
It explains, for instance,  the fact that the idea of viewing the super Poisson sigma model in the language 
of $Q$-bundles works well (cf. subsections \ref{sec:gpsm}, \ref{sec:gcs} below).

Let us understand what can be used as a target of the AKSZ in the multigraded case. 
Since the key difference to the ordinary approach consists in decoupling the 
$\Z_2$ parity from the $\Z$ ghost grading, we can consider all possible supersymplectic structures
on a multigraded manifold.
We can formulate the following ``classification'' result which is a straightforward 
generalization of \cite{Roytenberg2002}. 
\begin{prop}
 Symplectic structures on degree $1$ bigraded manifolds are in one-to-one correspondence with
super Poisson structures, on degree $2$ bigraded manifolds -- with Courant algebroids over 
a super manifold. 
\end{prop}

\begin{ex}
We can thus provide some natural examples 
 of gauge theories fitting into the
bigraded AKSZ construction described above: \emph{Poisson sigma model} for the target being a supermanifold with a Poisson structure, 
and \emph{Chern--Simons theory} with the target being a super Lie algebra (a relatively simple particular case of Courant algebroid).
\end{ex}
These models are particularly interesting since they include the description of 
supergravity in $1+1$ and $1+2$ space-time dimensions. 
For the analysis of the super-Poisson case and also some details of the super-Chern--Simons case
see the subsections \ref{sec:gpsm} and \ref{sec:gcs}.

\begin{rem}
The usual application of the AKSZ procedure, which is
is to provide a ``shortcut'' to construction 
of the Batalin-Vilkovisky action, can be thus implemented for these theories 
\end{rem}



\subsection{Source: measure on multigraded manifolds}

For the \textbf{super source} there is however a serious obstruction, namely 
the measure which is  homogeneous, non-degenerate and  invariant 
with respect to a given $Q$-structure does not always exist. 
More precisely the 
following proposition holds true:
\begin{prop}
  If the (non-negatively) multigraded manifold $M$ admits a non-degenerate homogeneous measure then its
 sheaf of functions contains only nilpotent or ghost degree 0 generators. 
 \end{prop}
\textbf{Proof.}  Suppose that a non-nilpotent element 
of positive ghost degree exists. 
As $\mu$ should be non-degenerate, the induced 
scalar product on $C^{\infty}(M)$ should be also non-degenerate. 
In particular, for any polynomial $p_k$ of degree $k$ in this element there 
exists a function $g_k$, such that $\int \mu p_k g_k \neq 0$.
Since $M$ is ghost $\Z_{\geq 0}$-graded and $p_k$ can be of arbitrary degree this is in contradiction 
with homogeneity of $\mu$.

For instance in the classical setting the usual example of a source manifold
is $T[1]\Sigma$ (for an ordinary manifold $\Sigma$) 
equipped with the $d_{de Rham}$ and ordinary berezinian measure. 
If we now replace $\Sigma$ by a supermanifold $\Sigma^{p|q}$, then
the super de Rham differential applied to odd coordinates of $\Sigma$
produces precisely the commuting elements of ghost degree $1$ forbidden
by the proposition. 

It would be interesting to understand the relation of the bigraded AKSZ construction
to the ordinary one, since the main motivation for applying it is the 
supersymmetrization of sigma model. 
Apparently considering possible bigraded source manifolds projecting to 
$T[1]\Sigma$ equipped with $d_{\text{de Rham}}$ when ignoring 
the super variables one does not have much choice: at least in the 
small degree case the associated Lie algebroid construction 
(\cite{Mackenzie}) results in something close to a direct product 
of $T[1]\Sigma \times R^{0|p}$ we will later comment on such examples.


There are however non-trivial cases when for a supermanifold a compatible pair $(Q, \mu)$ exists. 

\begin{ex} 
Let us consider $M = \mathbb{R}^{0|n}$ -- a purely odd vector space 
equipped with a $Q$-structure.
This corresponds to an $n$-dimensional Lie algebra structure 
(via the Chevalley-Eilenberg differential).
The standard berezinian measure on $M$ is compatible with a given $Q$-structure
if and only if $H^n_{Q}({\cal G},\mathbb{R}) = \mathbb{R}$.  
Indeed, if one views $Q$ as the map defining the cochain complex, 
vanishing of $\int Q f$ for any $f$ is equivalent to the image 
of $Q \colon \Lambda^{n-1} {\cal G}^* \to \Lambda^{n} {\cal G}^*$
being trivial. As for the reason of dimensions $Q$ always maps 
$\Lambda^{n} {\cal G}^*$ to $\{0\}$, that is $Ker Q_{|\Lambda^{n} {\cal G}^*} = \mathbb{R}$, 
it is equivalent to $H^n({\cal G},\mathbb{R}) = \mathbb{R}$. 
One can notice that the $H^n({\cal G},\mathbb{R}) = \mathbb{R}$
is related to ${\cal G}$ being a nilpotent Lie algebra.
If ${\cal G}$ is nilpotent, the cohomology is 
of dimension one, i.e. not vanishing. 
 
\end{ex}

In general the existence of the admissible measure should be of some relation to modular class of the $Q$-manifold (\cite{roytenberg_cirm}).

A natural idea is to consider the source manifold
being a product: something like $T[1]\Sigma \times \mathbb{R}^{0|p}$, 
or more general 
$(N_1 \times N_2, Q + Q', \mu \otimes \mu')$ in order to ``transfer''
the second factor to the target. Then the degrees should be compatible, i.e. 
for a $QP$-manifold $M, Q_2, \omega$ the degree of a symplectic form on 
the space $Maps(N_2, M)$ should be $gh(\omega) + gh(\mu_2)$.
For arbitrary product source manifolds 
the AKSZ construction on maps $N_1 \times N_2 \to M$ should give an equivalent 
result as on $N_1 \to Maps(N_2, M)$ if it is defined for both of them.
We will see some examples when for $Q' = 0$ it leads to on-shell equivalent theory,
but this will be better formalized in terms of section \ref{sec:qhom}.

\section{Multigraded Chern-Weil construction}
\label{sec:superCW}
This section is devoted to description of  $Q$-bundles that appear naturally 
in the context of gauging. Formally, the approach introduced in \cite{AK-TS}
is more powerful than the AKSZ procedure, in the sense that the latter permits to reproduce the former in 
a particular case, and also permits to treat more general theories not fitting into the AKSZ scheme.
We will however not profit much from this fact and rather view 
two approaches as complimentary ones, that is stressing different properties of the studied models.

\subsection{``Usual'' formalism}
The strategy  is to encode the fields of a gauge theory in a degree preserving map $\varphi$ between the $Q$-manifolds: 
source (world-sheet) (${\cal M}_1, Q_1$) and target $({\cal M}_2, Q_2)$.
In general a degree preserving map $\varphi$ between two $Q$-manifolds fails to be a $Q$-morphism, 
that is ${\cal F} = Q_1 \varphi^* - \varphi^* Q_2 \neq 0$.
But we can cure the situation by lifting the picture to the shifted tangent bundle of the manifolds 
${\cal M}_1$ and ${\cal M}_2$ and defining the map 
$ \varphi_* Q_1 - Q_2 \varphi$, $f : {\cal M}_1 \to T[1]{\cal M}_2$ covering $\varphi$.
$$
\xymatrix{
T[1]{\cal M}_1   \ar [rr]^{\varphi_*}       &&  T[1]{\cal M}_2 \\
&&&\\
\ar [uu]^{Q_1} {\cal M}_1 \ar [uurr]^{f} \ar [rr]^{\varphi}&& {\cal M}_2 \ar [uu]^{Q_2}& 
}
$$
Then $f^*$ is a $Q$-morphism between the $Q$-manifolds (${\cal M}_1, Q_1$) and 
$(\tilde{\cal M}, \tilde Q) = (T[1]{\cal M}_2, \rd _{DR} + {\cal L}_{Q_2})$,
where $\rd_ {DR}$ is the de Rham differential  on ${\cal M}_2$. 
Moreover, one can trivially extend $\varphi$, and thus $f$ to a mapping
${\cal M}_1 \to {\cal M}_1\times {\cal M}_2$, and 
${\cal M}_1 \to {\cal M}_1\times \tilde{\cal M} = \hat{\cal M}$ respectively (we denote them by the same letters), 
$\hat{\cal M}$ is equipped with the $Q$-structure $\hat Q = Q_1 + \tilde Q$.
This gives the following diagram
\begin{equation} \label{diag} \nonumber 
\xymatrix{
      && ( {\cal M}_1 \times \tilde {\cal M}) \circlearrowleft \hat Q \ar@/^/ [ddll]^{pr_1} \ar [dd]  \\
  \\
 {Q_1 \circlearrowright\cal M}_1 \ar@/^/[uurr]^{f} \ar [rr]^{\varphi}&& {\cal M}_1 \times {\cal M}_2 & 
} 
\end{equation}
The key idea is to consider the $Q$-bundle defined by the projection $pr_1$ --
gauge transformations can be then parametrized by 
$\hat \varepsilon$ -- vector fields on $\hat {\cal M}$
of total degree $-1$, vertical with respect to  $pr_1$:
\begin{equation} \label{gt}
 \delta_{\varepsilon}  ( f^*\cdot) = 
   f^*(  [\hat Q, \hat \varepsilon] \cdot).   
\end{equation}

This simply-looking equation (\ref{gt}) has interesting consequences. For example it permits to 
give a geometric interpretation of gauge symmetries for important theories, 
like the Dirac sigma model (DSM) defined in \cite{DSM} and studied in \cite{VS-TS}. 
One can also interpret the gauge invariance in terms of equivariant $\tilde Q$-cohomology 
of $\tilde {\cal M}$, and thus replace the procedure of gauging by searching for an
equivariantly $\tilde Q$-closed extension of a given superfunction: in \cite{VS-TS} the DSM is obtained in such a way, 
and details about the twisted Poisson sigma model (\cite{HPSM}) can be found in \cite{VSjgp}.

\begin{rem}
  A natural question to ask is how generic is the situation when one can proceed with the 
  above construction. For the world-sheet manifold usually there is no problem: one often considers 
  $T[1]\Sigma$ as the $Q$-manifold (cf. example \ref{ex:1}). For the target
  according to \cite{melchior} the $Q$-structure exists when the dimension of the 
  world-sheet is not too small, and 
  field equations satisfy 
  a certain type of Bianchi identities, which is not a very restrictive condition. 
  For the world-sheet dimension equal to two it has been shown in \cite{AK-VS-TS}, that 
  whenever gauging of the Wess-Zumino term is not obstructed, one recovers, modulo eventual degeneracies,
  a ``small'' Dirac structure, and thus  a $Q$-structure associated to it.
\end{rem}

\begin{rem}
  In the examples that interest us, the functional is the integral over the world-sheet
of a pull-back by $f^*$ of some natural object on the target. 
In the case of the Poisson sigma model, this object is just a symplectic form, in this sense 
one can recover the AKSZ formalism.
\end{rem}

\subsection{Multigraded generalization}  \label{sec:qhom}
We clearly see, that combining the previous subsection with the definitions of multigraded 
$Q$-manifolds from section \ref{sec:Gman} one can already formulate the generalization 
of \cite{AK-TS} in a rather straightforward way, and there is no need to spell this out.
Let us however stress one more important 
idea in the context. 
 
One of the messages of  \cite{BKS} (that somehow preceded \cite{AK-TS}) 
is that one can view the solutions 
of the field equations of the Poisson sigma model as morphisms of Lie algebroids
and the gauge transformations can be identified with the Lie algebroid homotopies. 
Since according to \cite{Vaintrob} Lie algebroids correspond to degree $1$ 
$Q$-manifolds, the statement about Lie algebroid morphisms is nothing but 
the idea that field equations correspond to $Q$-morphisms.
But the other part of the statement actually gives more information 
about the sigma model, since it permits to give a conceptual geometric explanation 
of  gauge transformations by associating them to $Q$-homotopies.

One can again formulate the multigraded version of the above observation.
Given a multigraded $Q$-manifold $(M,Q)$ the manifold $(M\times T[1]I,Q+\rd_I)$ is also a 
multigraded $Q$ manifold, therefore let us give  the following definition:
\begin{dfn}
  Two $Q$-morphisms $\varphi_0$ and $\varphi_1$
  between (multigraded) $Q$-manifolds $(M_1, Q_1)$ and $(M_2, Q_2)$ are called \emph{$Q$-homotopic} 
  if for $I = [0,1]$ there exists a $Q$-morphism $\varphi \colon (M_1\times T[1]I, Q_1 + \rd_I) 
  \to (M_2, Q_2)$ of (multigraded) $Q$-manifolds,
  the restriction of which to the boundary components 
  $M_1\times \{0\}$ and $M_1\times \{1\}$
  coincides with $\varphi_0$ and $\varphi_1$ respectively.
\end{dfn}
Going through the proof of the proposition in \cite{BKS} one sees
that the gauge transformations generated by a commutator 
of the total $Q$-structure with a degree $-1$ vertical vector field on the total target 
correspond to $Q$-homotopies.  
This observation is important, since it permits to study the space of solutions of the gauge theory 
not writing explicitly the functional, namely having encoded the source and the target 
geometries into $Q$-structures, consider the space of 
solutions modulo gauge transformations which is identified to 
$Q$-morphisms modulo $Q$-homotopies. 
This approach is useful since sometimes in 
physical applications the functional is not known or ill defined, 
but the equations of motion still have a geometric interpretation.

\section{Application to supersymmetric theories}
\label{sec:susy}

We are now going to apply the developed formalism to analysis of some supersymmetric 
theories. Our main examples will include the graded Poisson sigma model and super Chern--Simons theory.
The choice of these examples is motivated by the fact that in the ordinary 
(not multigraded) case both of them fit into the AKSZ scheme.
We will see what happens if the source and/or the the target are supersymmetrized, mathematically 
it means that smooth manifolds are replaced by supermanifolds.

\subsection{Super-target PSM}\label{sec:gpsm}

First we recover the generalized super version (\cite{1999}, \cite{2000}) of the Poisson sigma model
(\cite{PS-TS}). 
Recall that in an ordinary setting the Poisson sigma model is a gauge theory where the 
target space is a Poisson manifold. The field content is the following: 
scalar fields $X^i : \Sigma \to M$, $1$-form (``vector'') fields: $A_i \in \Omega^1(\Sigma, X^* T^* M)$. 
The action functional: 
$$  S = \int_{\Sigma} A_i \wedge \rd X^i + \frac{1}{2}\pi^{ij}A_i \wedge A_j, $$
where $\pi^{ij}$ are the components of the Poisson bivector on $M$.
One can equivalently say, that it is the functional on vector bundle morphisms from $T\Sigma$
to $T^*M$, this will give rise to degree preserving maps in the construction below.

In the super version $M$ carries also a ${\mathbb Z}_2$ grading, that is some of the fields 
$X^i, A_i$ are Grassmann valued. The Poisson bivector is now also non-trivially graded.
Actually the two super versions are not equivalent in the graded case: 
in \cite{1999} $S = \int A_i \rd X^i - \frac{1}{2}A_iA_j\pi^{ji}$, in \cite{2000}
$S = \int \rd X^i A_i + \frac{1}{2}\pi^{ji}A_iA_j$. 
But both of
them reduce (up to a total sign) to the usual Poisson Sigma model if all the fields are Grassmann-even. 

\textbf{Definition}

Let us note, that written as above the super Poisson sigma model is not well defined in the sense that the 
super (non-trivially $\mathbb{Z}_2$-graded) fields are parametrized by variables on a real (purely Grassmann-even)
manifold $\Sigma$. 
The first point to clarify is the definition of the ``base'' maps $X^i \in Map(\Sigma, M)$, where $M$ 
is a supermanifold. The standard way to cure this problem is the approach of $P$-points: 
one adds a parameter space -- a supermanifold $P$ to the source manifold.
Now the maps from $P\times\Sigma$ to $M$ are well-defined, as both the source and the target are
in the same category of supermanifolds. Then we can view $\Sigma$ itself as a supermanifold 
(by chance purely even, but it doesn't make any difference) and use the fact that 
$Hom(P\times\Sigma, M) = Hom(P, \underline{Hom}(\Sigma, M))$ (see \cite{Voronov} and the appendix \ref{sec:super_appendix} for details).
This equality should be considered as the equality of sets for all $P$ running over the 
category of supermanifolds. The left-hand-side of it 
is an honest set and it defines implicitly the space of maps, that interests us:  
$X^i \in \underline{Hom}(\Sigma, M)$. 
The next step is to define the ``one-form'' fields $A_i$. 
The procedure is essentially the same as before, 
one just has to 
remember, that the pull-back of bundles $X^*T^*M$ is also well-defined.

So we see that although the construction is rather complicated, formally one 
can just forget the parameter part of the source and view the fields as 
scalars and forms on $\Sigma$ with appropriate commutation relations, depending also 
on the super-grading of the variables on $M$. 
The same thing is true 
for the integrand of the functional $S$ which is always Grassmann-even: one just 
has to be careful about the order of the factors, because as can be seen from 
the two examples of \cite{1999} and \cite{2000}, different orderings produce 
non-equivalent theories. If one wants 
to define the integral itself one considers the Berezinian measure on the 
parameter space, but in fact it also doesn't influence much the ``physical'' 
part of the theory, meaning that the integration along an ordinary manifold $\Sigma$
can still be performed formally. 
The only thing to have in mind is that the result doesn't have to be a real number as 
in the ordinary case, but can be Grassmann valued (cf. also the definition of partial and full measures) 
that is for example,
$S = \int_{\Sigma} A_i \wedge \rd X^i - \frac{1}{2} A_i A_j \pi^{ij}  \in C^{\infty}(P)$.

\textbf{Sign conventions}

We are going to discuss the case of 
(\cite{2000}) and briefly comment on (\cite{1999}). In both articles the Deligne sign convention 
is used to 
synchronize the differential form degree ($\in\mathbb{Z}$) with the super grading 
($\in\mathbb{Z}_2$), 
we will also comment on the Bernstein-Leites sign convention 
in this context. As mentioned in section \ref{sec:Gman}, due to the isomorphism between 
two sign conventions one can even consider mixing them for multiple gradings, so this is not a 
conceptual choice.

So, first turn to the \emph{Deligne} sign convention.
In the $Q$-language the target manifold will be ${\cal M}_2 = T^*[1]M$ - the cotangent bundle to $M$
with the ${\mathbb Z}$ degree of the fiber variables shifted by $1$. For a super function on ${\cal M}_2$ 
there is a well defined 
notion of a super hamiltonian vector field, related to the canonical even symplectic form 
$\omega = \rd x^i \rd p_i$ 
$$
X_{f} = \{f, \} = (-1)^{|f||i|}\frac{\partial f}{\partial p_i} \frac{\partial}{\partial x^i} 
 - (-1)^{(|f| + 1)|i|}\frac{\partial f}{\partial x^i} \frac{\partial}{\partial p_i}
$$ 
where 
$|f|$ is the ${\mathbb Z}_2$ parity of $f$.
It permits us to construct a natural $Q$-structure on ${\cal M}_2$: 
$Q = X_{\Pi}$ for the hamiltonian superfunction $\Pi = \frac{1}{2} p_i \pi^{ij} p_j$. 
Let us note that the order of terms in $\Pi$ as well as the order of indeces in $\pi^{ij}$ is important, as 
in the super case all these terms are non-trivially $\Z_2$-graded: $|{ \pi^{ij} }|= |i| + |j|$, 
$\pi^{ij} = (-1)^{|i| |j| + 1} \pi^{ji}$.
$|\Pi| = 0$, so in local coordinates 
$$
  Q = \pi^{ij} p_j \frac{\partial }{\partial x^i}  + \frac{1}{2} \pi^{jk}_{,i} p_k p_j \frac{\partial }{\partial p_i} 
$$
The condition of $Q$ being homological is equivalent to the super Jacobi identity for the Poisson 
bracket on ${M}$, that is vanishing of the Schouten bracket of $\pi$ with itself.
Having $Q$ one performs the lifting procedure described in section \ref{sec:superCW}
and obtains
the $Q$-morphism $f$, that should now preserve separately both gradings: ${\mathbb Z}$, coming from the 
shifts of fibers in the bundles, and ${\mathbb Z}_2$, coming from Grassmann parity of the fields, 
that is with the same remark as before we permit odd functions on a purely even manifold $\Sigma$.

Defined like this, $f$ possesses some nice properties permitting to describe 
the ingredients of the sigma model. For example, for the gauge fields 
$A^\alpha = f^*(q^\alpha)$, $q^\alpha$ being any canonical coordinate on the target ${\cal M}_2$,
the field strengths will be just 
$F^\alpha = f^*(d q^\alpha)$, so one recovers naturally the equations of motion of the sigma model 
obtained in (\cite{2000}) in 
the form $f^*(d q^\alpha) = 0$:
$$
  \rd X^i + \pi^{ij}A_j = 0, \qquad \rd A_i + \frac{1}{2}\pi^{jk}_{,i}A_kA_j = 0
$$
As well as the gauge transformations:
$$
  \delta_{\varepsilon} X^i = \pi^{ij}\varepsilon_j, 
    \qquad \delta_{\varepsilon}A_i = -d{\varepsilon} - \frac{1}{2}\pi^{jk}_{,i}\varepsilon_kA_j
$$
where $\varepsilon_i$ are the components of a vector field 
$\varepsilon = \varepsilon_i \frac{\partial}{\partial p_i}$ on ${\cal M}_2$, 
and $\tilde \varepsilon$, lifted then to $T[1]{\cal M}_2$.
Moreover, one can easily see, that the integrand of the functional 
$$
  S_{gPSM} = \int_{\Sigma} \rd X^i A_i + \frac{1}{2}\pi^{ij}A_j \wedge A_i,
$$
could be obtained by partial integration of the pull-back of the symplectic form 
$\omega = \rd x^i \rd p_i $, that was used to define the hamiltonian vector field.

As for the second case (\cite{1999}), let us only give the starting point of the construction.
The main difference in comparison to the previous picture is that now it is more convenient
to consider left derivatives instead of the right ones, that is consider the vector fields on ${\cal M}$
as a right module and differential forms as a left module. Certainly, it creates some changes 
in the form of the hamiltonian vector fields on ${\cal M}$ but modulo this, one can just repeat the 
procedure taking the symplectic form $\omega = \rd p_i \rd x^i$ and the hamiltonian 
$\Pi = \frac{1}{2} p_i p_j \pi^{ji}$, and immediately recover the 
sigma model ingredients from (\cite{1999}).n

Let us note that in the picture described above we actually had to deal with three gradings: 
a $\Z_2$ Grassmann parity of the coordinates on $M$, and two $\Z$-gradings 
coming from the shifts of the gradings in $T[1]T^*[1]M$. In order to make 
the $Q-bundle$ construction work, the sign convention 
between the latter two has to be the Bernstein--Leites one. This is not in 
contradiction to any of the multigraded constructions defined above: 
at worst one has to go through the mentioned 
isomorphism  for each commutation relation to define the total sign.
But it is still slightly 
artificial to use the Deligne one for the $\Z_2$-parity. 
So below we describe the supersymmetric version of the PSM with the \emph{Bernstein--Leites} sign convention for the 
double grading. 

We again consider the target manifold ${\cal M} = T^*[1]M$.
Now a super hamiltonian vector field will have the form:
$$
X_{f} = \{f, \} = (-1)^{|f|(|i|+1)}\frac{\partial f}{\partial p_i} \frac{\partial}{\partial x^i} 
 + (-1)^{(|f| + 1)(|i|+1)}\frac{\partial f}{\partial x^i} \frac{\partial}{\partial p_i}
$$
Taking the hamiltonian superfunction $\Pi = \frac{1}{2} p_i \pi^{ij} p_j$
we construct a $Q$-structure on ${\cal M}$ 
$$
  Q =X_{\Pi} = \pi^{ij} p_j \frac{\partial }{\partial x^i}  + 
(-1)^{|i| + |j|}\frac{1}{2} \pi^{jk}_{,i} p_k p_j \frac{\partial }{\partial p_i} 
$$
Here $|\pi^{ij}| = |i| + |j|$, 
$\pi^{ij} = (-1)^{(|i|+1)(|j| + 1)} \pi^{ji}$.
$|\Pi| = 0$. The condition of $Q$ squaring to $0$ again amounts to the Jacobi identity. 

With the same procedure as before we obtain the equations of motion:
$$
  \rd X^i + \pi^{ij}A_j = 0, \qquad \rd A_i + (-1)^{|i| + |j|}\frac{1}{2}\pi^{jk}_{,i}A_kA_j = 0
$$
As well as the gauge transformations:
$$
  \delta_{\varepsilon} X^i = \pi^{ij}\varepsilon_j, 
    \qquad \delta_{\varepsilon}A_i = -\rd {\varepsilon} - (-1)^{|i| + |j|}\frac{1}{2}\pi^{jk}_{,i}\varepsilon_kA_j
$$
And pulling back the symplectic form $\omega = dp_i dx^i$ we obtain the functional
$$
  S_{gPSM} = \int_{\Sigma} \rd X^i A_i - \frac{1}{2}\pi^{ij}A_i \wedge A_j
$$
As before, this picture also obviously reduces 
to the ordinary one when all the variables are Grassmann even.


\subsection{Super-target Chern-Simons theory} \label{sec:gcs}
We are now going to apply the strategy of $Q$-bundles to 
the supersymmetric Chern-Simons theory. In contrast to the ordinary case, described 
as an example in \cite{AK-TS}, we consider the target manifold being a super Lie algebra
${\cal G}$, that is its generators $\xi^a$ will carry a ${\mathbb Z}_2$ grading (we will denote it 
$|a|$).

To follow the logic, discussed above 
one needs to consider the target space as a $Q$-manifold.
The natural way to do it is to take the ${\mathbb Z}$ graded shift of it with the Chevalley-Eilenberg differential as a 
$Q$-structure (cf. example \ref{ex:2}).
I.e. $({\cal M}, Q) = ({\cal G}[1],d_{CE})$, where local coordinates on ${\cal M}$ are $\xi^a$ with a double grading
$(|a|, 1) \in {\mathbb Z}_2\times {\mathbb Z}$. 
As we have discussed in the previous subsection it is more natural to consider 
the Bernstein-Leites sign convention for treating the multiple gradings, 
so we will present the computations only using it, noting that 
in the ``non-super'' 
(i.e. $|a| = 0, \forall a$) case the model reduces to the ordinary theory.
$$
  Q = \rd_{CE} = \xi^b \xi^cC^a_{bc} \frac{\partial}{\partial \xi^a}
$$
$Q^2 = 0$ is equivalent to the (Super) Jacobi identity for ${\cal G}$.
Equations of motion have the form:
$$
  F = \rd A + [A,A]_{gr} = 0,
$$
where $[\cdot, \cdot]_{gr}$ denotes the commutator coherent with the 
bracket defined for multigraded vector spaces and the chosen sign convention. 
And gauge transformations:
$$
  \delta_{\varepsilon} A^a = \rd \varepsilon^a + \varepsilon^b \xi^cC^a_{bc}
$$
The functional  reads
$$
  S = \int_{\Sigma} A \rd A +  < A, [A\wedge A]_{gr}>, 
$$
where again we should make a remark about integrating odd variables over an even manifold.


\subsection{Supersymmetrizing the world-sheet of the PSM leads to on-shell equivalent theory}
We now turn to analysis of supersymmetric source (world-sheet) manifold, that 
is instead of $\Sigma^2$ we consider explicitly $\Sigma^{2|m}$ - a supermanifold with $m$ Grassmann 
variables. As mentioned before, properly defining the functional
in this setting is difficult, so we 
study the space of $Q$-morphisms modulo $Q$-homotopies
between the multigraded manifolds appearing naturally in the context.
Certainly, extending $\Sigma$ produces extra degrees of freedom to our model, but they also produce 
more gauge symmetries. 
It turns out, that the  following proposition holds true:
\begin{prop}
The theory resulting from the source supersymmetrization of the PSM 
is on-shell equivalent to the original (non-supersymmetric) one.
 \end{prop}
To prove the proposition, we want to show, that the degrees of freedom are not physical, 
that is one can remove them on-shell by appropriate gauge fixing.
Let us introduce local coordinates on $\Sigma^{2,m}$: $\sigma^\mu$ for the even variables, and 
$\sigma^a$ for odd. Then, one can expand the fields with respect to Grassmann variables:
$$
a^*(x^I) = X^I = X^I_0(\sigma^\mu) + \tilde X^I(\sigma^\mu, \sigma^a), \qquad
a^*(p_I) = A_I = A_I^0 + \tilde A_I, \qquad
\varepsilon = \varepsilon_0 + \tilde \varepsilon
$$
where the index $I$ runs over all the variables on $M$, including the Grassmann-odd.
This expansion also changes other ingredients of the sigma model, i.e. all the functions 
of $X$ should now be considered as superfunctions. For example: 
$\pi^{IJ}(X) = \{X^I,X^J\} = \pi^{IJ}(X_0) + \pi^{IJ}_{,K} \tilde X^K  [+ \ldots]$.
We are interested in infinitesimal super variations of $X$, so one can consider all the expansions 
up to second order terms in $\tilde X$.  

Saying, that the super variations can be removed by gauge fixing, we mean that one can find 
$\tilde \varepsilon$, s.t. $\tilde X^I = \delta_{\tilde \varepsilon} X^I$ and 
$\tilde A_I = \delta_{\tilde \varepsilon} A_I$ -- this would precisely give the desired $Q$-homotopy.
The first condition means, that 
$\tilde X^I = \pi^{IJ} \tilde \varepsilon_J$, and is easy to satisfy. 
To see that let us consider two extreme cases: when $\pi^\#$ is invertible and when it is 
absolutely degenerate, i.e. vanishes.
In the first case we just take $\tilde \varepsilon_J = (\pi^{IJ})^{-1} \tilde X^I$, while in the second
$\tilde \varepsilon_J = 0$ and the $\tilde X^I$ cancels because of the equations of motion, having the 
form $\rd X^I = 0$.
The general case is the combination of two, that is we use the trivial $\tilde \varepsilon$ on the kernel of 
$\pi$ and the appropriate construction on the complement to it. 
The only problem we can face is in the neighborhood of the singularities of $\pi$, where we can not define these two subspaces. In this case one needs a more subtle argument like, that the differential $\rd$ respects the image of $\pi$, that is because of the structure of the equations of motion for $X$ we always have a solution $\tilde \varepsilon$.

Having constructed $\tilde \varepsilon$ one needs to check, that it satisfies also 
$-\tilde A_I = \rd \tilde \varepsilon + \pi^{JK}_{,I} \tilde \varepsilon_K A_J$.
The proof of it is direct computation, using the equations of motion, as well as the explicit expansion 
of $\pi(X_0 + \tilde X)$ and Jacobi identity for it, not forgetting, that we consider infinitesimal gauge 
transformations, i.e. all the equations are up to second order terms in `` $\widetilde{ }$ '' variables.

\begin{rem}
  This equivalence is easy to understand in the case of the most simple Poisson sigma model, 
  when $\pi = 0$. The integrand then reduces to the expression of the form 
  $A_I dX^I$, the field equations become the conditions of closedness of  $1$-forms
  and gauging consist of modifying the $1$-form field by the exact $1$-form. 
  That is the space of solutions modulo gauge transformations is described 
  the the corresponding cohomology. 
  But since the cohomology of a supermanifold coincides with the one of its body manifold 
  addition of super degrees of freedom extends the space of solutions by homotopies. 
  This is coherent with the on-shell equivalence result.   
\end{rem}


\subsection{Supersymmetrizing the world-sheet of Chern-Simons}
In this subsection we consider the multigraded manifolds in the context of
world-sheet supersymmetrized version of the Chern-Simons theory.
The resulting theory can 
be again formulated in the language of $Q$-manifolds.

Let us consider the world-sheet manifold as being the product of two $Q$-manifolds 
$({\cal M}_0 \times {\cal M}_1, Q_0 + Q_1)$ and the target manifold $({\cal M}_2, Q_2)$. We want to 
``transfer'' the second factor of the product to the target. To get a reasonable theory one 
needs to consider a new target manifold ${{\cal M}_2}^{{\cal M}_1}$ - a space of maps from ${\cal M}_1$ to 
${\cal M}_2$. There is a natural way to define a $Q$-structure on that target, therefore one can proceed 
with the construction. 

In the example that interests us the initial world-sheet can locally be viewed as 
$T[1](\Sigma \times \Pi {\mathbb R}^m)$ with the trivial $Q$-structure on the second factor. Then 
the resulting target turns out to be finite dimensional 
${\cal M} \equiv ({\cal G}\otimes_{\mathbb R} \wedge {\mathbb R}^m)[1] = (T[1])^m {\cal G} $. This is actually a (super) Lie algebra, with 
the bracket $[\eta_1 u_1, \eta_2 u_2] = \eta_1 \eta_2 [u_1, u_2]$, that is one can define a natural $Q$-structure
on it $Q = D_{CE}$ the Chevalley-Eilenberg differential on the total target, then the construction works as before, 
but on the extended target space. Therefore, we have proven the following proposition:
\begin{prop}
 The source supersymmetrized Chern--Simons theory can be reformulated as the target-supersymmetrized theory 
 with an extended algebra. 
\end{prop}
The question of equivalence of the theories thus reduces to analysis of super Lie algebras.

\subsection{AKSZ in supersymmetrization}

In this subsection we perform a sort of ``by hand'' supersymmetrization of some 
particular examples of the Poisson sigma model with small number of odd 
coordinates on the source. Those fit precisely to the multigraded AKSZ procedure, 
namely to the case of product manifolds we mentioned in the end of section \ref{sec:superaksz}.


\textbf{PSM (1,0)-SUSY}.

Consider \emph{one} odd coordinate on the world-sheet, i.e.
$\tilde {\cal M}_1 = {\cal M}_1 \times {\cal M}_2 = 
T[1]\Sigma \times \mathbb{R}^{0|1}$ with local coordinates 
$ \vartheta^{\mu}(1,0), \sigma^{\mu}(0,0), \theta(0,1)$,
numbers in brackets denote respectively the $\mathbb{Z}$-grading 
occurring from the shift of the fiber coordinates of the bundle $T[1]\Sigma$
and $\mathbb{Z}_2$-parity coming from the superextension.  
The target is the usual one ${\cal M}_3 = T^*[1]M$ for $M$ being an ordinary manifold, 
with local coordinates:
$p_i(1,0), x^i(0,0)$, the first $\mathbb{Z}$-grading in the brackets
comes from the shift of the fibers of the cotangent bundle, the second
trivial $\mathbb{Z}_2$-parity is induced by the source parity.

For the fields consider the space 
$Ho\underline{m}(\tilde{\cal M}_1, {\cal M}_3)$, 
the notation $Ho\underline{m}$ means that only the $\mathbb{Z}$-grading is preserved. 
That is the scalar fields have the form $X^i (0,0) = X^i_0(\sigma) + \theta X^i_1(\sigma)$
and the $1$-form valued fields: $A_i(1,0) = A^0_{i\mu}(\sigma)\vartheta^\mu + \theta A^1_{i\mu}(\sigma)\vartheta^\mu
=: A^0_{i} + \theta A^1_{i}$

With the standard Berezin integration on $\tilde{\cal M}_1$ the \emph{odd} functional 
has the form 
\begin{equation} \label{AdX_1}
  S = \int \rd\theta \int A_i\rd X^i = \int A_i^0\rd X^i_1 + A_i^1\rd X^i_0.
\end{equation}
Or more generally in the presence of the Poisson bivector
\begin{eqnarray} \label{SPSM_1}
  S = \int \rd\theta \int A_i\rd X^i + \frac12\pi^{ij}A_iA_j = \nonumber \\
= \int A_i^0\rd X^i_1 + A_i^1\rd X^i_0
+ \frac12\pi^{ij}(X_0)A_i^0A_j^1 + \frac12\pi^{ij}(X_0)A_i^1A_j^0 + \frac12\pi^{ij}_{,k}(X_0)X^k_1A_i^0A_j^0
\end{eqnarray}

We can now use the multigraded AKSZ construction 
to identify 
$$
Ho\underline{m}(\tilde{\cal M}_1, {\cal M}_3) \simeq 
Hom({\cal M}_1, Ho\underline{m}({\cal M}_2, {\cal M}_3))
$$ 
equipped with the appropriate 
structures. Namely, for our choice of 
${\cal M}_2 = \mathbb{R}^{0|1}$ and ${\cal M}_3 = T^*[1]M$, 
$Ho\underline{m}({\cal M}_2, {\cal M}_3) \simeq \Pi T(T^*[1]M) $ where $\Pi$ stands for the 
$\mathbb{Z}_2$-parity shift of the fiber coordinates of the tangent bundle, with 
local coordinates:
$ \psi_i(1,1), v^i(0,1), p_i(1,0), x^i(0,0)$. The AKSZ construction defines on this space of 
maps a symplectic form $\rd p_i\rd v^i + \rd\psi^i\rd x^i$, that is we identify 
$\Pi T(T^*[1]M) \simeq \Pi T^*[1]\Pi T M$, this permits to recover (using the Stokes' theorem)
the functional (\ref{AdX_1}).

If we now consider the case of the Poisson sigma model then on $T^*[1]M$ there is a non-trivial 
$Q$-structure 
$Q_0 = p_i\pi^{ij}\frac{\partial }{\partial x^j} + 
\frac{1}{2}p_ip_j\pi^{ij}_{,k}\frac{\partial }{\partial p_k}$
giving rise (again by multigraded AKSZ construction) to a $Q$-structure on $\Pi T^*[1]\Pi T M$:
\begin{eqnarray}
  Q_{ext} = 
    \theta \left( (\psi_i\pi^{ij} \pm \frac12p_i\pi^{ij}_{,k}v^k)\frac{\partial }{\partial x^j}
  + (\frac{1}{2}\psi_ip_j\pi^{ij}_{,k} + \frac{1}{2}p_i\psi_j\pi^{ij}_{,k} + 
     \frac{1}{2}p_ip_j\pi^{ij}_{,kl}v^l )\frac{\partial }{\partial p_k} \right. \nonumber \\
 \left. +  (p_i\pi^{ij})\frac{\partial }{\partial v^j} + 
    (\frac{1}{2}p_ip_j\pi^{ij}_{,k})\frac{\partial }{\partial \psi_k} \right) \nonumber
\end{eqnarray}
This $Q$ structure together with the symplectic form mentioned above permits to 
recover the functional (\ref{SPSM_1}).
To simplify the computations we can notice, that 
$Q_{ext} = \theta {\cal L}_{Q_0}$. Also one can check that only 
the antisymmetry and the Jacobi identity for the original Poisson bivector is needed for 
this operation (especially when using the Stokes' theorem). 
This computation is certainly trivial in the case of vanishing 
$\pi$. 

\textbf{PSM (1,1)-SUSY}.

Let us follow the same scheme for ${\cal M}_2 = \mathbb{R}^{0|2}$, i.e.
\emph{two} odd coordinates $\theta^{\nu}(0,1)$. Similarly, for the fields consider
$Ho\underline{m}(\tilde{\cal M}_1, {\cal M}_3)$.
$$
X^i (0,0) = X^i_0 + X^i_\nu\theta^{\nu} + X^i_2 \theta^1 \theta^2, \quad
A_i (1,0) = A_{i0} + A_{i\nu}\theta^{\nu} + A_{i2} \theta^1 \theta^2
$$
Integrating with the Berezinian the functional reads 
\begin{equation} \label{AdX_2}
  S = \int \rd\theta^1\rd\theta^2 \int A_i\rd X^i = \int A_i^0\rd X^i_2 + A_{i2}\rd X^i_0 + A_{i[\nu}\rd X^i_{\mu]}.
\end{equation}
Or more generally in the presence of the Poisson bivector
\begin{eqnarray} \label{SPSM_2}
  S = \int \rd\theta^1\rd\theta^2 \int A_i\rd X^i + \frac12\pi^{ij}A_iA_j  
= \int A_i^0\rd X^i_2 + A_{i2}\rd X^i_0 + A_{i[\nu}\rd X^i_{\mu]} + \nonumber \\
+ \frac12\pi^{ij}(X_0)A_{i0}A_{j2} + \frac12\pi^{ij}(X_0)A_{i2}A_{j0} + 
  \frac12\pi^{ij}(X_0)A_{i[\mu}A_{j\nu]} - \nonumber \\
- \pi^{ij}_{,k}(X_0)X^k_{[\mu}A_{i0}A_{j\nu]}
  +\left(\frac12\pi^{ij}_{,k}X^k_2 - \frac14\pi^{ij}_{,kl}X^k_{[\mu}X^l_{\nu]}\right)A_{i0}A_{j0} 
\end{eqnarray}

Let us consider the gauge transformations of the functional (\ref{AdX_2}).
$$
  \delta_\varepsilon X^i = 0, \quad \delta_\varepsilon A_i = \rd \varepsilon_i
$$
The same thing works for $\pi \neq 0$, but symmetries will
be dependent:
$$
\delta_\varepsilon X^i = \varepsilon_j\pi^{ji}(X), \quad 
\delta_\varepsilon A_i = \rd \varepsilon_i + \pi^{jk}_{,i}(X)A_j\varepsilon_k
$$

One can obtain the functionals (\ref{AdX_2}, \ref{SPSM_2}) by the AKSZ construction, 
more precisely again use the fact that 
$Ho\underline{m}(\mathbb{R}^{0|1}, {\cal M}) \simeq \Pi T({\cal M})$.
One should just be careful while iterating this procedure for 
$\mathbb{R}^{0|2} = \mathbb{R}^{0|1}\times \mathbb{R}^{0|1}$.
The space of maps is defined by
$$
\underline{Hom}(\mathbb{R}^{0|2}, T^*[1]M) \simeq \Pi T\Pi T(T^*[1]M)
\simeq T^*[1]\Pi T\Pi TM 
$$
For the second equality the identification works as follows.
\newline The coordinates on $\Pi T\Pi T(T^*[1]M)$: 
$$\psi'_i(1,1,1), v'^i(0,1,1), p_i'(1,0,1), x'^i(0,0,1), \psi_i(1,1,0), v^i(0,1,0), p_i(1,0,0), x^i(0,0,0)$$
on $T^*[1]\Pi T\Pi TM$: 
$$\bar v'_i(1,1,1), \bar x'_i(1,0,1), \bar v_i(1,1,0), \bar x_i(1,0,0), v'^i(0,1,1), x'^i(0,0,1), v^i(0,1,0), x^i(0,0,0)$$
To make things more transparent we write in brackets independently the parities that come from two 
$\mathbb{Z}_2$-parity shifts of fibers, the total $\mathbb{Z}_2$-parity is the sum modulo $2$ of 
the second and the third number in the brackets. Now identify the coordinates 
following the rule:
$q(0, a, b) \leftrightarrow q(0,a,b), \, q(1, a, b) \leftrightarrow q(1,a+1,b+1)$.
That is the odd symplectic form is given by $\omega = \rd x^i \rd\psi'_i + \rd v^i \rd p'_i + \rd x'^i \rd\psi_i + \rd v'^i \rd p_i$.
And for $\pi\neq0$ the even Poisson structure corresponds to the double lift of the vector
field $Q$.

Thus, with the multigraded AKSZ construction we are able to construct  
supersymmetrizations for a certain class of Poisson sigma models, namely to those 
defined on the tangent bundle to a Poisson manifold.

\section{Conclusions / discussion}

We have seen in this paper, that supersymmetric sigma models can be successfully 
formulated within the framework of multigraded geometry. 
One approach (generalization of the AKSZ procedure) permits to define functionals 
governed by symplectic forms on the target manifolds. But even 
if not all the ingredients of the AKSZ construction are present the other 
(much more general) approach of $Q$-morphisms and $Q$-homotopies permits to study the space of 
solutions for the physical theory and establish some equivalence results.

It would be however interesting to find non-trivial examples when 
the generalized AKSZ procedure produces a source supersymmetric 
theory that is not equivalent to any target supersymmetric theory. 
We have noticed that one of the main issues in working with multigraded 
AKSZ procedure is the existence of invariant measure on the source manifold. 
Several recent works may be useful in the context. 
First, in \cite{KwPo} some details of differential calculus on $\Z_2^n$-graded 
manifolds are studied -- among others, the integration theory closely resembling the 
computation of residues is established. 
Second, one is tempted to relax the compatibility conditions on the admissible objects in the AKSZ 
scheme. For example in \cite{Voronov_AKSZ} the \emph{difference construction} is supposed to 
replace the classical approach, a density on the source manifold is still part of the construction, 
but it does not look that crucial. Another conceptual approach is related to 
derived algebraic geometry: the construction in \cite{PTVV, CPTVV} includes the classical AKSZ
and may give some insight on the multigraded version.

It can also be fruitful to apply the generalized equivariant cohomology 
to multigraded manifolds with the motivation to recover and extend some 
characteristic classes appearing naturally in the context 
of supersymmetric gauge theories (cf. the series of papers \cite{niemi}).
Another issue which would be interesting to look at in the same spirit, 
is related to holomorphic analogs of the mentioned constructions. 
On one hand in the recent years physicists were dealing with the likewise constructions in 
the context of sigma models (cf. the papers \cite{holsm}),
on the other hand, mathematicians were interested in it in the context 
of higher structures replacing the smooth manifolds by algebraic, holomorphic, 
K\"ahler etc (cf. \cite{UBVR}, \cite{VRetal}). It would be good to couple these two subjects 
using graded geometry. \\[2em]

\textbf{Acknowledgements.}
I am thankful to Thomas Strobl for constant attention towards this activity. 
I would like to thank Jean-Philippe Michel, who was the driving force in the beginning 
of this project.
I also greatly appreciate inspiring discussions with Alexei Kotov, Valentin Ovsienko,   
 Vladimir Roubtsov,  Dmitry Roytenberg, Florian Sch\"atz, Theodore Voronov at 
various stages of this work. \\
My current research is supported by the Fonds National de la Recherche, 
Luxembourg, project F1R-MTH-AFR-080000.

\newpage

\appendix

\section{Super and graded geometry}
\label{sec:super_appendix}

This appendix is given here for the sake of completeness of the paper and also to fix some (standard)
notations. We give a recollection of results from super/graded geometry following \cite{leites, bern_sem, GR9, GR11, Roytenberg2002, Roytenberg}.

\subsection{Supermanifolds}

\begin{dfn}
  A supermanifold $M$ is a ringed space $(M_0, {\cal O}_M)$, where ${\cal O}_M$ is a sheaf 
  of commutative superalgebras  on $M_0$, such that $M_0$ is a Hausdorff topological 
  space with a countable base, and every point $m\in M_0$ has a neighborhood $U_0$, 
  such that the ringed space $(U_0, {\cal O}_M |_U)$ is isomorphic to a superdomain $U = (U_0, {\cal O}_U)$.
\end{dfn}
A \textit{morphism of supermanifolds} $\varphi \colon M \to N$
is a morphism of the corresponding ringed spaces. The set of morphisms will be denoted 
$Hom(M, N)$. A morphism $\varphi \colon M \to N$ is called a diffeomorphism
if there exists an inverse morphism $\varphi^{-1} \colon N \to M$.
The global subsections of a sheaf ${\cal O}_M$ are called \textit{superfunctions} on $M$.
For a morphism $\varphi \colon M \to N$, one denotes $\varphi_0 \colon M_0 \to N_0$
the corresponding map of body manifolds, $\varphi^* \colon C^{\infty}(N) \to C^{\infty}(M)$
the corresponding morphism of superalgebras (or sheaves ${\cal O}_N  \to {\cal O}_M$). 
The following theorem holds:

\begin{theo} \label{th:batch} (Batchelor -- Gawedzki, \cite{batchelor, gawedzki})
For any supermanifold $(M, {\cal O}_M)$ there exists a vector 
bundle $E \to M_0$ such that $(M, {\cal O}) \simeq \Pi E$ (non-canonically isomorphic).
\end{theo}

The points of the body $M_0$ of a supermanifold 
are not enough to define all the structure of it, there is however a way to formalize the usual intuition 
of viewing a manifold as a set of points, namely so called parameter spaces.
The idea (\cite{leites}) is that we can consider the families 
of properties depending also on the odd parameters. 
For supermanifolds $P, M, N$ a $P$-family of morphisms $\varphi_P \colon M \to N$
is a morphism $\varphi \colon P \times M \to N$. We can also consider an equivalent 
construction $\varphi' \colon P \times M \to P \times N$.
Any morphism of supermanifolds $\varphi \colon M \to N$ can be considered as a $pt$-family of
morphisms, where\footnote{``A point is a point -- there is no point in this'', -- P.~ \u{S}evera, talk at Poisson 2016 Conference.} $pt = \R^{0|0}$.
A $P$-family of points of a supermanifold $M$ is a $P$-family of morphisms 
$pt^P \colon pt \to M$, i.e. a morphism $pt \colon P \to M$ such a morphism
is usually called simply a \emph{$P$-point of a supermanifold $M$}.
Although a supermanifold is not defined by all its points it is 
defined by its $P$-points for all supermanifolds $P$.
This concept is important for definition of some objects in the context 
of supersymmetric sigma models, where we will mention it explicitly. 
 


\subsection{Graded geometry, $Q$-manifolds}

\subsubsection{Graded manifolds} \label{sec:gman}

\begin{dfn}
  A graded vector space $V$ is a collection of vector spaces $ V = \oplus V_i, \, i\in \Z$ or 
  $i\in \Z_{\geq 0}$. 
\end{dfn}
For an element $v_i \in V_i$ we shall denote its grading by $gh(v_i) = i$ or $|v_i| = i$.
The notation $gh$ is chosen since the physical meaning that is often 
attributed to the $\Z$-grading of a superfunction is the so-called
\emph{ghost number}.

Given two graded vector spaces $V$ and $W$ it is natural to define
a graded space $\underline{Hom}(V,W)$ of \textit{homomorphisms} from $V$ to $W$.
The homomorphism $\varphi \colon V \to W$ is $p$-graded ($\varphi \in \underline{Hom}_{p}(V, W)$)
if it maps $V_i \to W_{i+p}$. The space $Hom(V, W) \equiv \underline{Hom}_{0}(V, W)$ will be called the space of
\textit{morphisms}
from $V$ to $W$.
There is a canonical homomorphism shifting the grading by $p$, denoted usually by $[p]$:
$(V[p])_{i} = V_{i-p}$.
Since one usually assumes the base field to be of degree $0$, the 
\textit{dual} vector space $(V_i)^*$ is defined as $(V^*)_{-i}$.

\begin{dfn}
  A graded algebra is a graded vector space $A$ with multiplication: the operation 
  $\cdot \colon A \otimes A \to A$, compatible with the grading, i.e. $\cdot \colon A_p \otimes A_q \to A_{p+q}$.
  A degree preserving homomorphism of algebras is called a morphism of graded algebras.
\end{dfn}
Since in what follows there will be sometimes several gradings in the construction,
to avoid confusion instead of writing ``morphism'' we will specify if a 
homomorphism preserves some grading. 

The direct analog of the \textit{sign rule} is applicable for the computations 
in the graded algebras: the sign $(-1)^{gh_1 gh_2}$ appears when a $gh_1$-graded element 
passes through a $gh_2$-graded element. Following this rule one defines 
the \textit{graded commutator} $[a,b] = ab - (-1)^{gh(a)gh(b)}ba $ and the 
\textit{graded commutative} elements, when $[a,b] = 0$.

The \textit{graded symmetric algebra} $S(V)$ over a graded vector space $M$ is the quotient 
of the tensor algebra over $V$ by the graded commutator ideal. 
That is $S(V)$ is spanned by the polynomials $f_{\nu_1 \dots \nu_k} \xi^{\nu_1}  \dots \xi^{\nu_k}$, 
where $\nu_i$ is a multiindex corresponding to the elements of $V_i$.

Fix a graded vector space $V$ and an ordinary manifold $M_0$. Then defining 
the \textit{graded manifold} $M$ morally 
is just extending the structure sheaf ${\cal O}_{M_0} = C^{\infty}(M_0)$ by $S(V)$. 
\begin{dfn} \label{grman}
 A graded manifold $M$ is a couple $(M_0, {\cal O}_M)$, where $M_0$ is a smooth manifold 
 and the sheaf of functions ${\cal O}_M$ is locally isomorphic to $C^{\infty}(U_0) \otimes S(V)$, 
 where $U_0$ is an open subset of $M_0$. 
\end{dfn}

For a vector space $V = V_1 \oplus \dots \oplus V_k$, 
$k$ is called \emph{degree} of $V$. 
Like for graded vector spaces, for graded manifolds the top degree of the generators of the 
structure sheaf is called \emph{degree} of a graded manifold. 

In contrast to the theorem (\ref{th:batch}) a graded manifold 
is not always locally described by a vector bundle. 
More precisely the following theorem holds for non-negatively 
graded manifolds.
\begin{theo} (\cite{Roytenberg2002}) Given a non-negatively graded 
manifold $(M, {\cal O}_M)$ there is a tower of fibrations 
$$
  M = M_n \to M_{n+1} \to \dots \to M_1 \to M_0,
$$
where any $M_k$ is a graded manifold of degree at most $k$, for $k>0$
$M_{k+1} \to M_k$ is an affine bundle.  
\end{theo}

In view of the above theorem it is easy to see that one can also describe 
the grading by introducing the Euler vector field 
$\epsilon = gh(x_{\alpha})\frac{\partial}{\partial x_{\alpha}}$.
According to \cite{GR9}, global definition of the Euler vector field on $M$
is equivalent to the definition of a smooth action of the monoid $\R_+$ on $M$.
More explicitly given a homogeneous coordinate system on $M$ 
(with the basis being the eigen directions of $\epsilon$) consider 
the action given by a homothety-type map $h \colon \R_+ \times M \to M$ such that 
$(x^1, \dots, x^N) \mapsto h_t(x^1, \dots, x^N) \equiv 
(t^{gh(x^1)}x^1, \dots, t^{gh(x^N)}x^N)$. 
In the converse direction the Euler vector field can be 
recovered as $\epsilon = \left. \frac{\partial }{\partial t} \right|_{t = 1} h_t$.

Such a map $h$ is called a \emph{homogeneity structure}.
In what follows we will actually prefer this equivalent definition
to (\ref{grman}), since defined like this the notion of a graded manifold permits a rather straightforward generalization 
  to multigraded. Morally one can choose a graded vector space $V$
  consider a supermanifold $M$ instead of $M_0$,    
  and define a graded manifold ${\cal M}$ over it by extending the sheaf of functions 
  ${\cal O}_M$ by $S(V)$. We introduce the appropriate notion in the main text.

Graded manifolds form a category ${\cal G}Man$ with $\underline{Hom}(M,N) = 
\underline{Hom}({\cal O}_N, {\cal O}_M)$.
One also has a forgetful functor into the category ${\cal S}Man$ of supermanifolds 
taking the degree modulo $2$. Note that in the category ${\cal S}Man$ the sheaf 
${\cal O}_M$ is extended by the smooth functions of the even degree variables\footnote{This is actually 
a rather deep analytic question that we address in \cite{JPS}}. 
\begin{prop} (\cite{Roytenberg}) \label{prop_gm}
   For fixed graded manifolds $M$ and $N$ the functor from ${\cal G}Man$ to ${\cal S}ets$
   given by $Z \to Hom(N\times Z, M)$ is \emph{representable}, 
   i.e. there exists a graded manifold $\underline{Hom}(N,M)$, such that $Hom(N\times Z,M) = Hom(Z,\underline{Hom}(N,M))$;
   its base $\underline{Hom}_0(N,M)$ is $Hom(N,M)$ viewed as an infinite dimensional smooth manifold 
   containing $\underline{Hom}(N_0,M_0)$.
\end{prop}
In this setting taking $N$ to be a point one recovers $M$ as $\underline{Hom}(pt,M)$.

\subsubsection{$Q$-manifolds}\label{sec:qman}
\begin{dfn}
A $Q$-manifold  (differential graded manifold)  is a graded 
manifold equipped with a degree $1$ vector field $Q$, which is homological, i.e.
self-(super)-commuting:
$$
  [Q,Q] =2 Q^2 = 0	
$$ 
\end{dfn}
An important result of \cite{Vaintrob} is that degree $1$ 
$Q$-manifolds are in one-to-one correspondence with Lie algebroids, therefore all the 
examples of Lie algebroids provide examples of $Q$-manifolds.
Let us be more explicit on this here. \newline
\begin{ex} \label{ex:1}
Consider the fiber-linear 
coordinates of the tangent bundle $T\Sigma$ of an ordinary (smooth) manifold $\Sigma$ as carrying degree plus one and 
those coming from the base as degree zero, we obtain a $\Z$-graded manifold, which usually is 
denoted by $T[1]\Sigma$. If we consider the grading only up to $2\Z$, we obtain a supermanifold, 
usually 
denoted by $\Pi T\Sigma$. Clearly  differential forms on $\Sigma$ become just functions on $T[1]\Sigma$ in this 
language, the form-degree mapping to the eigenvalue of the respective function with respect to the 
Euler vector field. The de Rham differential on $\Sigma$ increases the form-degree by one and squares to 
zero. It thus defines a $Q$-structure on $T[1]\Sigma$. 
In local coordinates $\sigma^\mu$ of degree $0$ and $\theta^\mu = \ud \sigma^\mu$of degree $1$ 
the $Q$-structure reads
$Q = \rd_{DR} = \theta^\mu \frac{\partial}{\partial \sigma^\mu}$. 
\end{ex}
\begin{ex} \label{ex:2}
Another important example of a $Q$-manifold is given 
by a Lie algebra: in fact any (odd) vector space $V$ shifted in $\Z$-degree by one which is equipped 
with a $Q$ vector field  is equivalent to the definition 
of a Lie algebra structure on $V$. 
If we choose the degree $1$ coordinates $\xi^a$ on  $V[1]$,
the $Q$-structure reads 
$d_{CE} = \frac12 C^{a}_{bc} \xi^b \xi^c \frac{\partial}{\partial \xi^a}$, 
$d_{CE}$ is a so-called Chevalley-Eilenberg 
differential, defining $C^a_{bc}$ -- the structure constants of the corresponding
Lie algebra ${\cal G}$. Permitting $C^a_{bc}$ to be non-constant functions 
on $M$ one recovers the action Lie algebroid. 
\end{ex}

Thus $Q$-manifolds permit a unified description of important geometric and algebraic 
structures. 
More involved examples of $Q$-manifolds can be constructed starting from 
 Poisson manifolds $M$, 
where the cotangent bundle $T^*[1]M$ carries such a vector field $Q$ canonically, 
or also by Courant algebroids -- we will discuss them later in the context of sigma models.

\subsubsection{$QP$, $NQP$-manifolds}

A non-negatively graded $Q$-manifold is sometimes called \emph{$NQ$-manifold}.

As for an ordinary manifold a \emph{symplectic form} is a closed non-degenerate $2$-form.
The grading of coordinates induces the grading of the symplectic form.

\begin{dfn}
  A \emph{$P$-structure} is a degree $1$ symplectic structure on a graded manifold.
\end{dfn}
A typical example is a $2$-form $\omega = \rd p_i \rd x^i$
of degree $1$
canonically associated to $T^*[1]M$; let us note that here 
saying that $\omega$ is of degree $1$, we distinguish the grading coming from the shift and the 
differential form degree in the total $\Z$-grading. 

A vector field $X$ is \emph{compatible} with the symplectic form 
if ${\cal L}_X \omega = 0$, where the usual notion of a Lie derivative extends to 
graded objects by a graded Cartan formula:
${\cal L}_X = \rd \iota_X + (-1)^{gh(X)} \iota_X \rd$.

\begin{dfn}
  A $QP$-manifold is a graded manifold with compatible $Q$ and $P$ structures. 
\end{dfn}

One can naturally classify the $NQP$-manifold for small degrees, namely the following 
statement holds true:
\begin{prop} (\cite{Roytenberg2002})
 Degree $1$ $QP$-manifolds are in one-to-one correspondence with Poisson manifolds; 
 degree $2$ $QP$-manifolds are in one-to-one correspondence with Courant algebroids. 
 \end{prop}

   \vspace{2cm}

\end{document}